\documentclass[apj]{emulateapj}
\usepackage{bm}
\usepackage{dcolumn}
\usepackage{graphicx}
\newif\ifAMStwofonts
\AMStwofontstrue 

\def\cl{{C_l}}

\def\dl{{d_{lm}}}

\def\tal{{{\tilde a}_{lm}}}
\def\dl{{d_{lm}}}

\def\gsim{~\rlap{$>$}{\lower 1.0ex\hbox{$\sim$}}}

\def\simpropto{\lower.2ex\hbox{$\; \buildrel \propto \over \sim \;$}}
\def\ltsim{\lower.5ex\hbox{$\; \buildrel < \over \sim \;$}}
\def\gtsim{\lower.5ex\hbox{$\; \buildrel > \over \sim \;$}}
\def\ltsim{\lower.5ex\hbox{$\; \buildrel < \over \sim \;$}}
\def\gtsim{\lower.5ex\hbox{$\; \buildrel > \over \sim \;$}}

\def\dd{\,{\rm d}}

\def\hmpc{\ {\rm h^{-1}Mpc}}

\def\dd{{\rm d}}

\def\ln{{\rm ln}}

\def\pmb#1{\setbox0=\hbox{#1}%
\kern-.025em\copy0\kern-\wd0
\kern.05em\copy0\kern-\wd0
\kern-.025em\raise.0433em\box0}

\def\vr{\pmb{$r$}}
\def\vn{\pmb{$n$}}
\def\hvn{\hat {\vr}}

\def\vk{\pmb{$k$}}

\def\simlt{\lower.5ex\hbox{$\; \buildrel < \over \sim \;$}}
\def\simgt{\lower.5ex\hbox{$\; \buildrel > \over \sim \;$}}

\newcommand{\beq}{\begin{equation}}
\newcommand{\eeq}{\end{equation}}
\def\beqa{\begin{eqnarray}}
\def\eeqa{\end{eqnarray}}
\def\fixit#1{}
\def\hmpc{h^{-1}\,{\rm Mpc}}

\def\dd{{\rm d}}

\def\cN{{\cal N}}

\shorttitle{Clustering  of radio galaxies}
\shortauthors{ Nusser  \& Tiwari}
\begin{document}
\title{The clustering of radio galaxies: biasing and   evolution versus  stellar mass}
\author{Adi Nusser }
\email{adi@physics.technion.ac.il}
\author{Prabhakar Tiwari}
\email{ptiwari@physics.technion.ac.il}
\affil{Physics Department and the Asher Space Science Institute-Technion, Haifa 32000, Israel\\
e-mail: adi@physics.technion.ac.il}
    
\date{\today}
\begin{abstract}

We study the angular clustering of  $\sim 6\times 10^5$ NVSS sources on  scales $\gtsim 50 \hmpc$ in the context of the $\Lambda$CDM scenario.  The analysis partially relies  on the redshift distribution of 131  radio galaxies, inferred from the Hercules and CENSORS survey,  and an empirical fit to the   stellar to  halo mass  (SHM) relation.
For redshifts $z\ltsim 0.7$, the  fraction of radio activity versus  stellar mass evolves as $f_{_{\rm RL}}\sim M_*^{\alpha_0+\alpha_1 z}$ where 
$\alpha_0=2.529{\pm0.184}  $  and 
 $\alpha_1=1.854^{+0.708}_{-0.761}$. The estimate on $\alpha_0$ is largely driven by the results of 
\cite{Best2005a},  while the constraint on $\alpha_1$ is new.  
 We derive a 
 biasing factor $b(z=0.5)=2.093^{+0.164}_{-0.109}$ between radio galaxies and the underlying mass. 
The function  $b(z)=0.33z^2+0.85 z +1.6$ fits well the redshift dependence.
We also provide convenient parametric forms for the redshift dependent radio luminosity function,  which are consistent with the redshift distribution and the NVSS source count versus  flux.

 \end{abstract}

\maketitle
\section{Introduction}
\label{sec:int}

Radio activity is a marker for energetic processes related to  star formation and 
accretion of gas on super massive black holes (SMBHs) at the nuclei of galaxies. 
Active Galactic Nuclei (AGN) are associated with cycles  of powerful jets generated by 
accretion disks on SMBH. These jets make a sizable contribution to the 
 energy budget 
at moderate redshifts and can play an important  role in galaxy formation
  \citep[e.g.][]{1995MNRAS.276..663B,Bower2006,Croton2006} and even offsetting gas cooling in 
massive groups and  clusters of galaxies \citep[e.g.][]{1993MNRAS.264L..25B,1994MNRAS.270..173C,McNamara2000}.
Therefore,  the interplay between   radio activity and  other galaxy characteristics  
is of great interest \citep[e.g][]{Best2005a,Fontanot2011}. 
The environment, whether 
galaxies are isolated or in dense regions, may also have an important impact on  the emergence of radio activity. 
Galaxy-galaxy interactions tend to boost  accretion of gas on the 
central SMBH \citep[e.g.][]{Best2005a,Pasquali2009}, eventually triggering AGN activity. 
One approach is  to directly assess   the observed galaxy environment of radio  AGNs 
\citep[e.g.][]{1988MNRAS.230..131P,Hill1991,McLure2001,Best2005a,Worpel2013}, but this is typically limited by small sample uncertainties. 
Alternatively,  the clustering of  radio galaxies   quantified through two-point statistics (e.g. correlation functions),   could constrain the way  these galaxies are related to the underlying mass density field and  and the properties of the host dark halos \citep[e.g.][]{1991MNRAS.253..307P,Lacy2000,Passmoor2013,Blake2004}. 
Unfortunately,  redshift surveys of radio galaxies contain a small number of objects\citep{Mauch2007,Brown2011,Donoso2010,VanVelzen2012}
and are currently unsuitable for a robust determination of two point statistics at moderate redshifts.
Here we revisit  the angular power spectrum computed from 
radio sources in the NRAO VLA Sky Survey (NVSS) \citep{NVSSC}. The survey provides angular positions 
and 1.4GHz fluxes, $S$, for about 80\% of the sky and is currently the most suitable survey for angular 
angular clustering  analysis.

Two  ingredients are needed for   extracting information from angular  clustering statistics  within 
a cosmological scenario,  e.g. the $\Lambda$CDM.
The first is  a model for the  mean redshift distribution, $N(z)$, of NVSS sources (galaxies). 
 We base our modeling of  $N(z)$ on two catalogs of radio sources with redshifts but small angular coverage.  These are the  Combined EIS-NVSS Survey  (CENSORS)\citep{Best2003,Rigby2011} 
and the  Hercules survey  \citep{Waddington2001} which are (jointly) complete down to a flux limit of $7.2\rm mJy$. 
The second ingredient is the biasing  relation  between the spatial distribution of   radio galaxies and the underlying mass density field.  Our modeling of this relation
adopts  a parametric form for the frequency of radio activity versus  stellar mass and
 an observationally motivated  stellar to halo mass (SHM) ratio from the literature. Well established 
halo biasing is then incorporated to yield a prescription for the  biasing relation of radio galaxies.  
 The angular power spectrum of radio galaxies as predicted in the 
  $\Lambda$CDM  is then contrasted with  the clustering of NVSS galaxies in order to put constraints the biasing relation and the prevalence of radio activity as a function of the stellar mass and redshift.

 The outline of the paper is as follows. 
 In \S\ref{sec:aps} we review the  definition of the angular power spectrum, $C_l$, and how it is 
 estimated from the data with partial sky coverage.  The calculation of the theoretical  $C_l$ in the context of a cosmological model 
  is presented in \S\ref{sec:tpk}. This section also 
 describes our modeling of the biasing relation and the redshift distribution of radio galaxies.
 In \S\ref{sec:like} we give the maximum likelihood methodology for parameter estimation. 
Constraints on the biasing and the evolution of radio activity are  derived in \S\ref{sec:results}.
In \S\ref{sec:consist}, our model for the redshift distribution is shown to be consistent with 
observations of the radio luminosity function and the source counts in the NVSS.
We end in \S\ref{sec:summary} with a general discussion of the results.  
\section{The angular power spectrum }

\label{sec:aps}
We  define the  surface number density contrast of sources
\begin{equation}
\label{eq:den}
\Delta (\hvn)=\frac{\cN (\hvn)}{\bar \cN}-1\; ,
\end{equation}
where $\cN(\hvn)$ is the projected number density (per steradian) in the direction $\hvn$ and $\bar \cN$ is the mean of 
$\cN$ over the sky. 
In practice  $\cN(\hvn)$ is obtained by dividing  the sky into small pixels (e.g. via HEALPix \citep{Gorski2005}) and counting the number of sources in each pixel.
%
%
%
We decompose $\Delta(\hvn)$ in spherical harmonics $Y_{lm}$ as  
\begin{equation}
\label{eq:flm}
\dl= \int_{\rm survey}   d \Omega  \Delta(\hvn) Y_{lm}(\hvn) \; .
\end{equation}
We imagine now that $\dl$ is one out of an ensemble of many  realizations of the same underlying 
continuous field.
The variance, $<|\dl|^2>_{_{\rm ens}}$, from this ensemble is approximated as \citep{Peeb80}, 
\begin{equation}
\label{eq:dlm}
<|\dl|^2>_{_{\rm ens}}=\left(C_l+\frac{1}{\bar \cN}\right)J_{lm}\; ,
\end{equation}
where the term involving $\bar \cN$ (mean number density over the observed part of the sky)  appears  due to the Poisson  discrete sampling (shot-noise) of the continuous field by a finite number of sources  and
\begin{equation}
J_{lm}=\int_{_{\rm survey}}|Y_{lm}|^2 \dd \Omega\; ,
\end{equation}
accounts for the partial sky coverage and  gives unity for  full sky data. The quantity $C_l$ is the underlying angular power spectrum in the 
continuous limit ($\bar \cN\rightarrow \infty$) for full sky coverage. It is independent of $m$ by the assumption 
of cosmological isotropy, i.e. the lack of a preferred direction.  According to (\ref{eq:dlm}),  an estimate of the true power spectrum from the observations is 
\begin{equation}
\label{eq:Cobs}
C^{\rm obs}_l=\frac{1}{2l+1}\sum_m\frac{|\dl|^2}{J_{lm}} -\frac{1}{\bar \cN}\; .
\end{equation}
The $1\sigma $ error in this estimate is \citep[][ cf.  \S\ref{sec:like} here]{Peeb80}
\begin{equation}
\label{eq:clsig}
\sigma_{_{\cl}}=\sqrt{\frac{2}{2l+1}}\left(\cl^{\rm obs}+\frac{1}{\bar \cN}\right)\; 
\end{equation}
reflecting cosmic variance and shot-noise scatter.
\subsection{The measured $C_l$}
\label{sec:mcl}
\begin{figure} 
\includegraphics[width=0.48\textwidth]{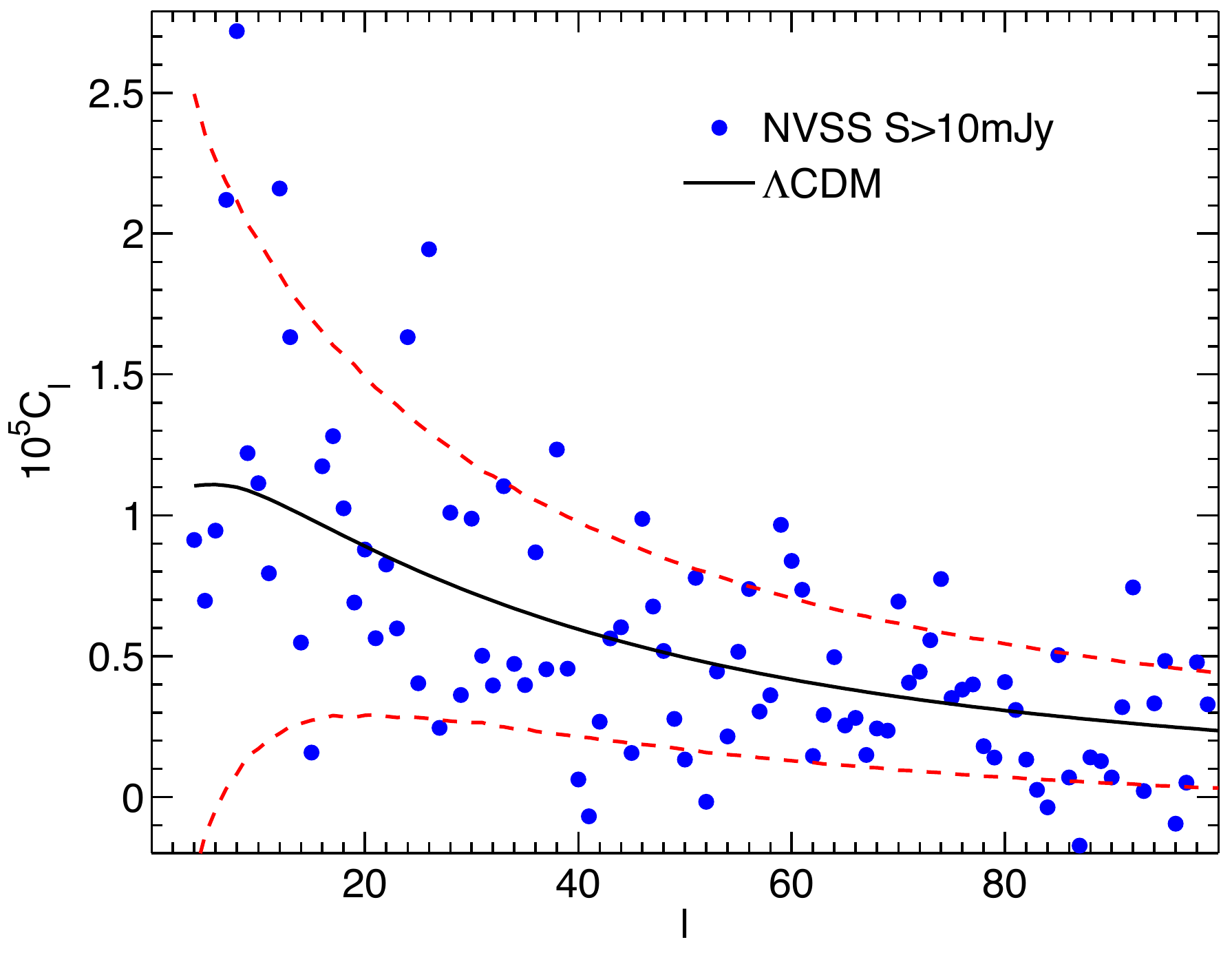}
\caption{The observed angular power spectrum estimated from the NVSS data for $S>10 \rm mJy$ is shown as the blue dots. 
The solid curve is the $\Lambda$CDM   ${\tilde C}_l$ obtained using  the ML parameters given in Table 1.
 The dashed red curves correspond to $1\sigma   $ limits due to shot-noise and cosmic variance scatter. }
 \label{fig:Cl}
\end{figure}

The NVSS 
contains $\sim$1.7 billion sources with angular positions at  1.4 GHz integrated  flux density $S>2.5\rm mJy$.
The full width at half maximum resolution is 45 arcsec and nearly all observations are
at uniform sensitivity.
The catalogue covers the sky north of declination $-40^\circ$ (J2000) which is almost
80$\%$ of the celestial sphere.  We trim the data as follows  \citep{Blake:2002}.
\begin{itemize}
 \item  Galactic sources at latitudes $|b|<5^{\circ }$
are masked out to avoid Galactic contamination. 
\item  We remove 22 bright extended local radio galaxies. 
\item  
As shown by \cite{Blake:2002},  the NVSS has significant
systematic gradients in surface density for  sources fainter than $10 ~\rm mJy$. Therefore,  the analysis is restricted to sources brighter than 10 mJy. 
We also discard extra
bright sources ($S>1000~\rm mJy$). 
\end{itemize}
This leaves us with 574466 sources which we use here  
to derive the  observed  angular power spectrum, $C^{\rm obs}_l$,
according to (\ref{eq:Cobs}).   The effect of multiple-component sources (e.g. radio lobes of the same galaxies) have been removed following the recipe of \cite{Blake2004}.
The blue filled dots 
 in figure \ref{fig:Cl} are $C^{\rm obs}_l$ for all values of $l$ between $l>l_{\rm min}=4$ and $l<l_{\rm max}=100$ corresponding
  to  comoving physical scales of   $\gtsim 50 \rm \hmpc$ at $z=0.5$.
The partial sky coverage introduces  mixing of power between different $l$-modes. 
We have assessed this mixing in random realizations of the surface density $\Delta$, generated from the $\Lambda$CDM cosmological model (see below) and found it to be negligible for $l>3$.  
Hence, we impose  a lower cut at  $l=l_{\rm min}=4$.  Since we shall restrict the analysis to large scales, we also impose a 
high $l$ threshold at  $l=l_{\rm max}=100$. In practice, for   $l>l_{\rm max}=100$, the signal to noise ratio becomes very small 
such that no significant information is lost by the restriction to $l<l_{lmax}$.
 We have confirmed that   $C^{\rm obs}_l$ in this figure agrees very well  with   \cite{Blake2004}.

\subsection{Theoretical  $C_l$}
\label{sec:tpk}

We derive here theoretical expressions for  angular power spectra of $\Delta(\hvn)$ 
 within a cosmological scenario. 
The redshift is denoted by $z$ and the comoving coordinate by $\vr$.
By $z(r) $ we mean the redshift of a particle at a comoving distance $r$ in a uniform cosmological background.  The underlying mass density field is $\rho^{\rm m}(\vr,z)$ and the corresponding 
 density contrast is  $\delta^{\rm m}(\vr,z)=\rho^{\rm m}(\vr,z)/\bar \rho^{\rm m}  -1$ where $\bar \rho^{\rm m}$ is the mean background density.  The density contrast of galaxies is denoted by $\delta$. 
 
 The theoretical counterpart of $\Delta(\hvn)$ is
\begin{equation}
\tilde \Delta (\hvn) =\int _{0}^{\infty} \delta(\hvn r, z(r)) p(r) \dd r\; ,
\end{equation}
where $\delta$ is the density contrast in the three dimensional distribution of galaxies at a comoving coordinate $\vr$ 
 redshift $z=z(r)$.
  The quantity $p(r) \dd r $
is the probability of observing a galaxy between   $r$ and  $(r+\dd r)$.

The  cosmological model provides the statistical properties 
 of the  underlying mass density contrast   field $\delta^{\rm m}(\vr,z)$, rather than the galaxy density contrast $\delta$.
 Restricting  the analysis to large scales (larger than a few 10s of Mpc)
we assume here a linear biasing relation 
 $\delta(\vr,z)=b(z) \delta^m(\vr,z)$,
as motivated by the measured three dimensional correlation function of  galaxies in redshift surveys \citep[e.g.][]{Norberg2001,Zehavi2011,Davis2011} and  analytic  studies of galaxy formation \citep{Nusser2014}.  
Further, on large scales,  linear gravitational instability paradigm yields  $\delta^{\rm m}(\vr,z)=\delta^m_0(\vr) D(z)$ where
$D$ is the linear growth factor of linear fluctuations normalized to unity at the present time, $z=0$   \citep{Peeb80}.
Therefore,  
\begin{eqnarray}
\tal&=&\int d \Omega \tilde \Delta Y_{lm}(\hvn)\\
\nonumber &=& \int d\Omega Y_{lm}(\hvn) \int_{0}^{\infty} W(r) \delta_0(\hvn r) d r\; ,
\end{eqnarray}
where $W=D(z)b(z)  p(r)$ with $z=z(r)$.
Going to the Fourier domain
\begin{equation}
\delta_0(\vr)=\frac{1}{(2\pi)^3}\int d^3 k \delta_{\vk}{\rm e}^{i \vk \cdot{  \vr}}\; ,
\end{equation}
and substituting 
$
\label{eq:ekr}
{\rm e}^{i \vk \cdot{  \vr}}=4\pi \sum_l {\rm i}^lj_l(kr) Y^*_{lm}(\hat {\vn})Y_{lm}(\hat {\vk})
$,
yields,
\begin{equation}
\tal=\frac{{\rm i}^l}{2\pi^2}\int d r W\int d^3 k \delta_{\vk}j_l(kr) Y^*_{lm}(\hat {\vk}) \; . 
\end{equation}
Therefore,
\begin{eqnarray}
\label{eq:clphi}
{\tilde C}_l&=&<|\tal |^2>\\
\nonumber &=& \frac{2}{\pi }\int dk k^2 P(k) \left\vert \int_{0}^{\infty} d r W j_l(kr)\right\vert^2 \; .
\end{eqnarray}
and we have used $<\delta_{\vk}\delta_{\vk'}>=(2\pi)^3 \delta^{\rm D}(\vk-\vk')P(k)$ where $\delta^{\rm D}$ is Dirac's $\delta-$function.
The power spectrum $P(k)$  of linear density fluctuations  is well constrained by a variety 
of observations, primarily the cosmic microwave anisotropies measured by the Planck satellite \citep{Planck2015}. 
The remaining ingredients  in the theoretical modeling are the radial distribution $p(r)$ and 
the biasing $b(z)$. 

\subsubsection{The radial  distribution} 

\begin{figure} 
\includegraphics[width=0.48\textwidth]{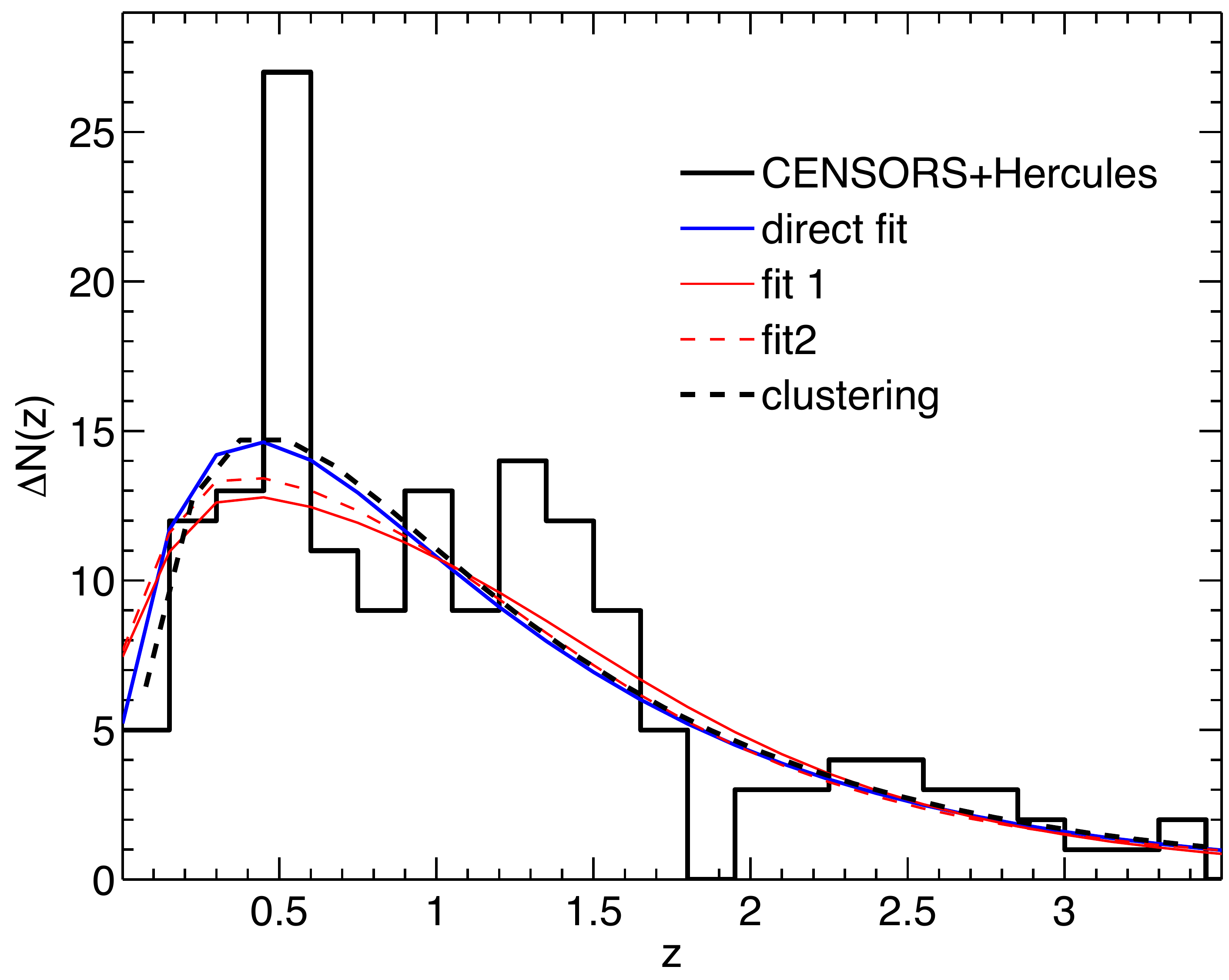}
\caption{ The histogram is the observed number of sources with $S>7.2~\rm mJy$  per  redshift bin $\Delta z=0.15$. The curves  ``fit~1" 
\& ``fit~2"  are obtained using two parametric forms for the radio luminosity function as described in \S\ref{sec:consist}.  }
 \label{fig:Nz}
\end{figure}

The modeling of   $p(r)$ relies on  CENSORS )\citep{Best2003,Rigby2011}   and Hercules \citep{Waddington2001} surveys of radio galaxies with observed  redshifts and  1.4GHz fluxes. 
CENSORS contains
 135 radio sources over $\rm 6 deg^2$ complete down to $7.2\; \rm mJy$   1.4GHz flux limit. Spectroscopic redshifts
are available for about $73\% $ of the sources. The remaining sources have been  assigned redshifts  through 
 the $I-z$ or $K-z$ magnitude-redshift relations.   
 In Hercules, there are  64 objects  1.2 $\rm deg^2$ with $S>2\; \rm mJy$.
The two surveys together contain 165 and 131 sources, respectively for  $S>\; \rm 7.2mJy$
and $10\; \rm mJy$. 

The number of sources per unit redshift, $N(z)$, in the two surveys jointly   is the basis for 
our model of $p(r)\propto N(z)\dd z/\dd r$. 
The distribution $N(z)$ is represented by the histogram  in figure \ref{fig:Nz}.
According to \cite{Rigby2011}, there is a difficulty in obtaining spectroscopic 
redshifts at  $z\approx 1.5$, resulting in the observed gap  in $N(z)$.
 Less accurate redshift measurements compensate but 
the random error scatter objects to the sides. As we shall see at a later stage,  the angular clustering probes
large scale structure $\ltsim 0.7$. Hence, 
the  results  are insensitive to  whether galaxies near the gap are included or not in the modeling of $p(r)$.
 Since the  number of galaxies is too small for a robust measurement of $N(z)$,  we adopt  the parametric  form for 
$ N(z)\propto p(r)  \dd r/\dd z$,
\begin{equation}
\label{eq:Nmodel}
N^{\rm model}\propto z^{a_1}\exp\left[ -\left(\frac{z}{a_2}\right)^{a_3} \right]\; ,
\end{equation}
where the parameters $a_1$, $a_2$ and $a_3$ are determined by maximizing the probability distribution 
for measuring  $N(z)$ as well as the  observed angular power spectrum, $C_l$. 
Further, 
we impose a upper redshift cutoff in the abundance of sources at $z=3.5$, consistent with decline in their number density 
as seen by \cite{Rigby2011}. In this modeling  we use the redshift distribution of galaxies with $ S>10\rm mJy$ to match the flux limit of the trimmed version of the  NVSS, as described in \S\ref{sec:mcl}. The normalization of $N^{\rm model}$ is such that it  yields the total number of sources ($S>10\rm mJy$) in the joint CENSORS and Hercules data. For a Poisson distribution of $N$, this is equivalent to performing a maginalization over the normalization.

\subsubsection{The biasing relation}

We will resort to  a variety of observational results  to link radio galaxies to the masses of their host dark matter  halos.
A description for the biasing of radio galaxies will then be obtained 
from  well-established relations between  the halo distribution and the mass density field.

 \cite{Best2005a}  studied the properties of the host galaxies  of a local  ($z<0.1$) sample of radio-loud AGN. They  demonstrated: $i)$ the presence of  radio activity is a strong function, $M_*^{2.5}$, of the stellar mass, $M_*$, of 
 the host galaxies, saturating at  $\sim 30\%$ for  $M_*\gtsim 10^{12}M_\odot$. 
$ii)$ the  shape of the radio luminosity distribution  in galaxies with a similar stellar mass, $M_*$,  depends very weakly on $M_*$ and is close to the global luminosity function  \citep{Mauch2007,Best2014a}.
We assume that similar  trends extend to higher redshifts and write the fraction 
 of radio loud AGNs with radio luminosity brighter than $P$ as
\begin{equation}
\label{eq:frlL}
f_{_{\rm RL}}={\cal F}(M_*,z) \Phi(P,z) \; ,
\end{equation}
where $ \Phi(P,0)$
 is close to the global radio luminosity. 
 Further, we adopt
 \begin{equation}
\label{eq:alpha}
{\cal F}(M_*,z)=\left(\frac{M_*}{10^{11}M_\odot }\right)^{\alpha_0+\alpha_1 z} \; ,
\end{equation} 
where  $\alpha_0=2.5\pm 0.2$ \citep{Best2005a} will be included as a prior in the maximum likelihood analysis presented below. The fraction of radio activity is practically
nonexistence at $M_*<10^{10}M_\odot$  \citep{Best2005a}. 
One of the main goals of this paper is  
to constrain  $\alpha_1$ from the clustering of the NVSS 
galaxies and the inferred $N(z)$ from the CENSORS and Hercules.
Since the stellar and halo masses are  closely related, the dependence of radio activity on $M_*$ dictates the 
biasing of the radio galaxies through the well studied halo biasing.  We  assume  that the radio luminosity depends on the halo mass, $M$, only through 
$M_*$  hence the form of $ \tilde \Phi(P,z)$ is irrelevant for  the biasing prescription.
We  further  assume that there is only radio galaxy per halo. This is incorrect for low luminosities as  seen in galaxy clusters \citep{Branchesi2005}, but it is a reasonable assumption for the luminosities considered here. 
Therefore, we write the biasing factor  as 
\begin{equation}
\label{eq:bias}
b(z)=\frac{\int_{M_{\rm min}}^{M_{\rm max}} \dd M n_{\rm h}(M,z)  b_{\rm h}(M,z){\cal F}(M_*,z)}{\int_{M_{\rm d}}^{M_{\rm u}} \dd M n_{\rm h}(M,z){\cal  F}(M_*,z)} \; ,
\end{equation}
where the stellar mass $M_*$ as a function of the halo mass, $M$, and redshift is taken from the SHM relation derived by 
\citep{Moster2013}. This  relation is  in reasonable agreement with a variety of observations at low and high redshifts
 \citep[c.f.][]{Coupon2015}. 
The mass thresholds  $M_{\rm min}$ and $M_{\rm max}$ are, respectively,   the minimum and maximum halo masses that can host a radio galaxy.
We take $M_{\rm min}=4\times 10^{11}M_\odot$, corresponding to  $M_*=10^{10}M_\odot$  \citep{Moster2013} which is 
the lowest  stellar mass exhibiting radio activity. As we will see later, because of the steep dependence of ${\cal F}$ on $M_*$, the results are insensitive to the exact value of $M_{\rm min}$. At the high mass end we take $M_{\rm max}=10^{15}M_\odot$, the mass scale of rich galaxy clusters. Halos with this large mass are rare, implying that our analysis in insensitive to this upper bound as well.

\section{Likelihood analysis}
\label{sec:like}
Within the context of the $\Lambda$CDM cosmology, we seek constraints on the 5  parameters 
$\alpha_0$, $\alpha_1$, $a_1$, $a_2$ and $a_2$ by maximizing 
the likelihood, $P_{\rm tot}$,  for observing $\cl^{\rm d}$ (from NVSS) and $N(z)$ (from the joint CENSORS and Hercules catalogs). 
For brevity, we denote these parameters  by $\theta_i$  ($i=1\cdots 5$). We derive now the expression for $P_{\rm tot}$.
Let  $\tilde d_{lm}=d_{lm}/J^{1/2}_{lm}$ and continue to  assume that  the partial sky coverage does not introduce 
significant  statistical correlation between  $\tilde d_{lm}$ with different $(l,m)$. Restricting the analysis to large scales, the probability distribution function (PDF)  of  $\tilde d_{lm} $ is gaussian   \begin{equation}
P(\tilde d_{lm})\propto \frac{1} {(\cl+1/ \bar \cN)^{1/2}}{\exp}\left[- \frac{1}{2} \frac{\tilde d_{lm}^2}{\cl+1/{\bar \cN}} \right]\; ,
\end{equation}
where  $<\tilde d_{lm}^2>=\cl+1/\bar \cN$ in accordance with  (\ref{eq:dlm}). 
The PDF of the full  set $\{\tilde d_{lm}\}$ of all $m=-l\cdots +l$  is
\begin{equation}
\label{eq:allm}
P_l\propto \frac{1} {(\cl+1/\bar \cN)^{l+1/2}}{\exp}\left(- \frac{2l+1}{2} \frac{\cl^{\rm d}}{\cl+1/\bar \cN} \right)\;  ,
\end{equation}
where 
 $\cl^{\rm d} =(\sum_m d_{lm}^2)/(2l+1)$.
Maximizing $P_l$   with respect to $\cl$ yields
$\cl=\cl^{\rm d}-{\bar \cN}^{-1}=C^{\rm obs}_l$ (c.f. eq.~\ref{eq:dlm}) as expected. An estimate of the $1\sigma $ error on this $\cl$ is
$\sigma_{_{\cl}}=(-\dd^2\ln P/\dd \cl^2)^{-1/2}=(\cl+{\bar \cN}^{-1})\sqrt{2/(2l+1)}$ (cf. Eq.~\ref{eq:clsig}).
We write the likelihood function for observing  $\cl^d$ and $N(z)$ as
\begin{equation}
\label{eq:Ptot}
P_{\rm tot}(\{\theta_i\};N,\cl^{\rm d})= P_{\alpha_0} \prod_{z_i} P_{_{\rm {N_i}}} \prod_{l=l_{\rm min}}^{l_{\rm max}}P_l \; ,
\end{equation}
where $ P_{\alpha_0}$ a (Gaussian) prior  PDF incorporating the   local, $z\approx 0$,  observational constraint   $\alpha_0=2.5\pm 0.2$  of \cite{Best2005a}.
Further,   $P_{_{\rm {N_i}}}$ is a Poisson PDF   for observing  $N_i$ galaxies in the redshift bin
$i$ given a mean number of  $N^{\rm model}(z_i) \Delta z$.  
We will use the redshift distribution of sources with  $S>10\rm mJy$ and $\Delta z=0.15$. 
The probability  $P_l$ is given by (\ref{eq:allm})  with $\cl=\tilde C_l$ computed using the  \cite{EH98} parametrized form of the 
 $\Lambda$CDM density power spectrum, $P(k)$. 
The cosmological parameters are taken from the latest Planck results \citep{Planck2015}:  Hubble constant $H_0=67.8 \rm km s^{-1} Mpc^{-1}$, total matter density parameter  
$\Omega_m=0.308$, baryonic density parameter $\Omega_b=0.0486$, linear clustering amplitude 
on $8\hmpc$ scale, $\sigma_8=0.815$ and a spectral index  $n_s=0.9667$.

The  evaluation of ${\tilde C}_l$   with full direct numerical integration is very slow. 
Fortunately,  the  Limber approximation 
is accurate to better than  
5\% even at small $l$, as revealed by comparison with the full calculation for some
relevant choices of the parameters.  
Substitution of the  Limber approximation  
\begin{equation}
\int d r W(r) j_l(kr )\approx \sqrt{\frac{\pi}{ 2l+1}} k^{-1}W\left(r=\frac{l+1/2}{k}\right)
\end{equation} 
in (\ref{eq:clphi}) yields, 
\begin{eqnarray}
\label{eq:limber}
\nonumber
{\tilde C}_l&=&\frac{2}{2l+1)}\int_0^\infty d k P(k) W^2\left(r=\frac{l+1/2}{k}\right)\\
&=&\int_0^\infty d r r^{-2} P\left(\frac{l+1/2}{r}\right) W^2(r)\; .
\end{eqnarray}

\section{Results} 
\label{sec:results}
We  fix  $l_{\rm min}=4$ and  $l_{\rm max}=100$, as the fiducial values.  
Table 1 lists the parameters  $\theta_i=\theta^{\rm ML}_i$ obtained by maximizing $P_{\rm tot}$.
The theoretical ${\tilde C}_l$  computed  with these parameters, is  shown in figure \ref{fig:Cl} as the black solid curve.
The dashed red lines are $1\sigma $ deviations from ${\tilde C}_l$ due to the combined error of shot-noise and cosmic variance, 
i.e., $\pm ({\tilde C}_l+1/\bar \cN)\sqrt{2/(2l+1)}$. 
The scatter of the blue points around ${\tilde C}_l$ is
entirely consistent with this error,  as confirmed using   random samples of $d_{lm}$ 
generated  assuming ${\tilde C}_l$ is the true power spectrum.
The redshift distribution $N^{\rm model}$ obtained with  ML parameters is plotted as the dotted black curve in figure 
\ref{fig:Nz}. This curve is remarkably close to the blue solid curve corresponding to a direct fit of $N^{\rm model}$ 
to the observed redshift distribution alone.

\begin{figure} 
\includegraphics[width=0.49\textwidth]{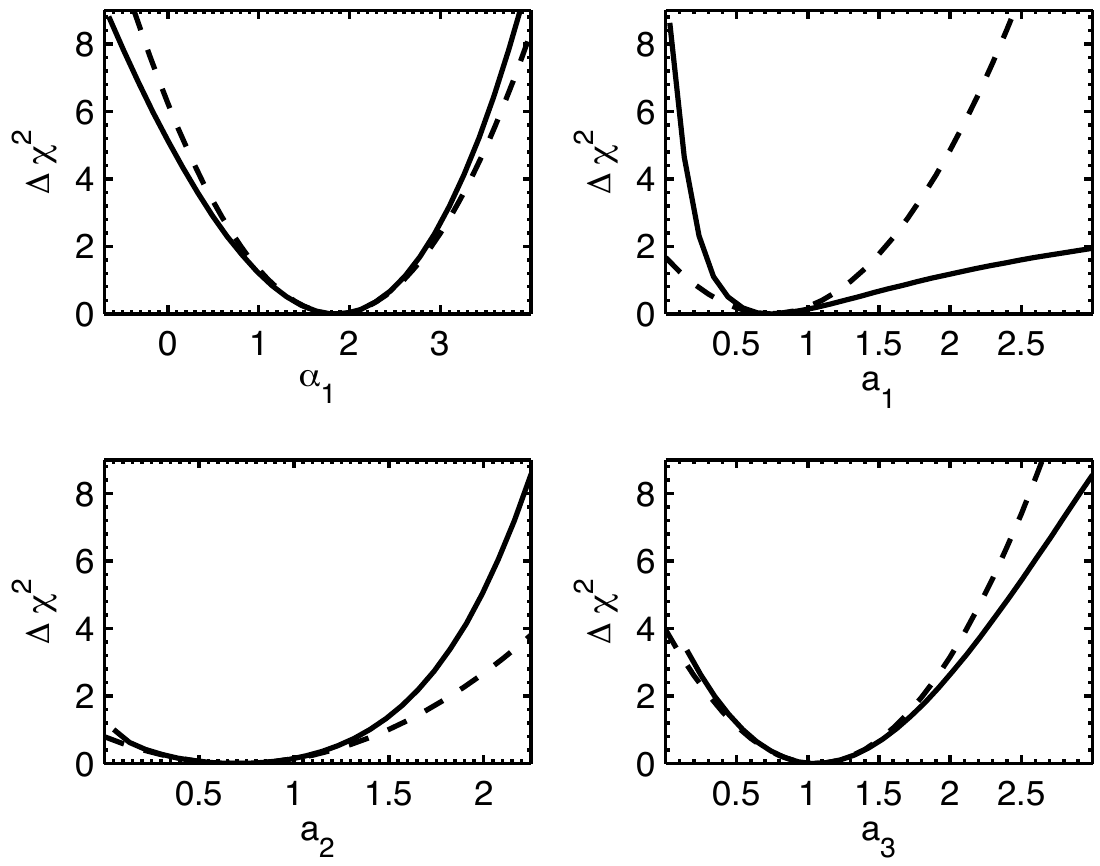}
\caption{ Confidence levels $ \Delta \chi^2=-2\ln (P_{\rm 1D}/P_{\rm 1D}^{\rm max})$ 
as a function of individual  fitting parameters (except $a_0$). In each panel, marginalization is performed over the remaining parameters, as described in the text.  
}
 \label{fig:cdf}
\end{figure}

Confidence levels on a single 
parameter $\theta_i$ are obtained by marginalization over the remaining 4  parameters as follows. For a given  $\theta_i$, we compute $\{\theta_j\}$ ($j\ne i$) which render a maximum in   
$P_{\rm tot}(\theta_i,\{\theta_{j\ne i}\})$. We denote by $P_{\rm 1D}(\theta_i)$, the value  $P_{\rm tot}$ evaluated 
at these $\{\theta_j\}$.  The confidence limit is then
$\Delta \chi^2(\theta_i) \equiv -2\ln (P_{\rm 1D}/P_{\rm 1D}^{\rm max})$ where $P_{\rm 1D}^{\rm max}$
is the maximum of $P_{\rm 1D}(\theta_i)$ which is also  the maximum  $P_{\rm tot}$
with respect to all 5 parameters.  
The solid curves in the 4 panels in figure \ref{fig:cdf}  represent   $\Delta \chi^2$ for $\alpha_1$, $a_1$, $a_2$ and $a_3$.
The dashed curve in each panel is an  approximation to   $P_{1D}$ by means of a  normal PDF. 
The corresponding figure for  $\alpha_0$  turns out to be highly symmetric with the solid and dashed 
curves almost completely overlapping and 
we opted  to omit it here.
The parameter, $\alpha_1$,  the indicator for the evolution of radio activity versus stellar mass,  is slightly skewed to the left relative to a normal distribution (dashed curve).
The  remaining parameters pertaining to $N^{\rm model}(z)$ are highly asymmetric. 
The super and sub scripts attached to the ML values in Table 1 represent the asymmetric $1\sigma$  errors
corresponding to  $\Delta \chi^2=1$ for the solid lines in the figure. 

The parameter $\alpha_0$ and the corresponding error in the Table are very close to the values used in the prior $P_{\alpha_0}$, indicating that $\alpha_0$ is mainly constrained by this prior rather than $N(z)$ and $C^{\rm obs}_l$.  
 Fitting  $N^{\rm model}(z)$ directly to the observed distribution alone without incorporating any information about clustering yields  best fit values  $a_1=1.199 $,   $a_2=0.323$,   $a_3= 0.81$ in agreement with the ML estimates.

\begin{figure} 
\includegraphics[width=0.49\textwidth]{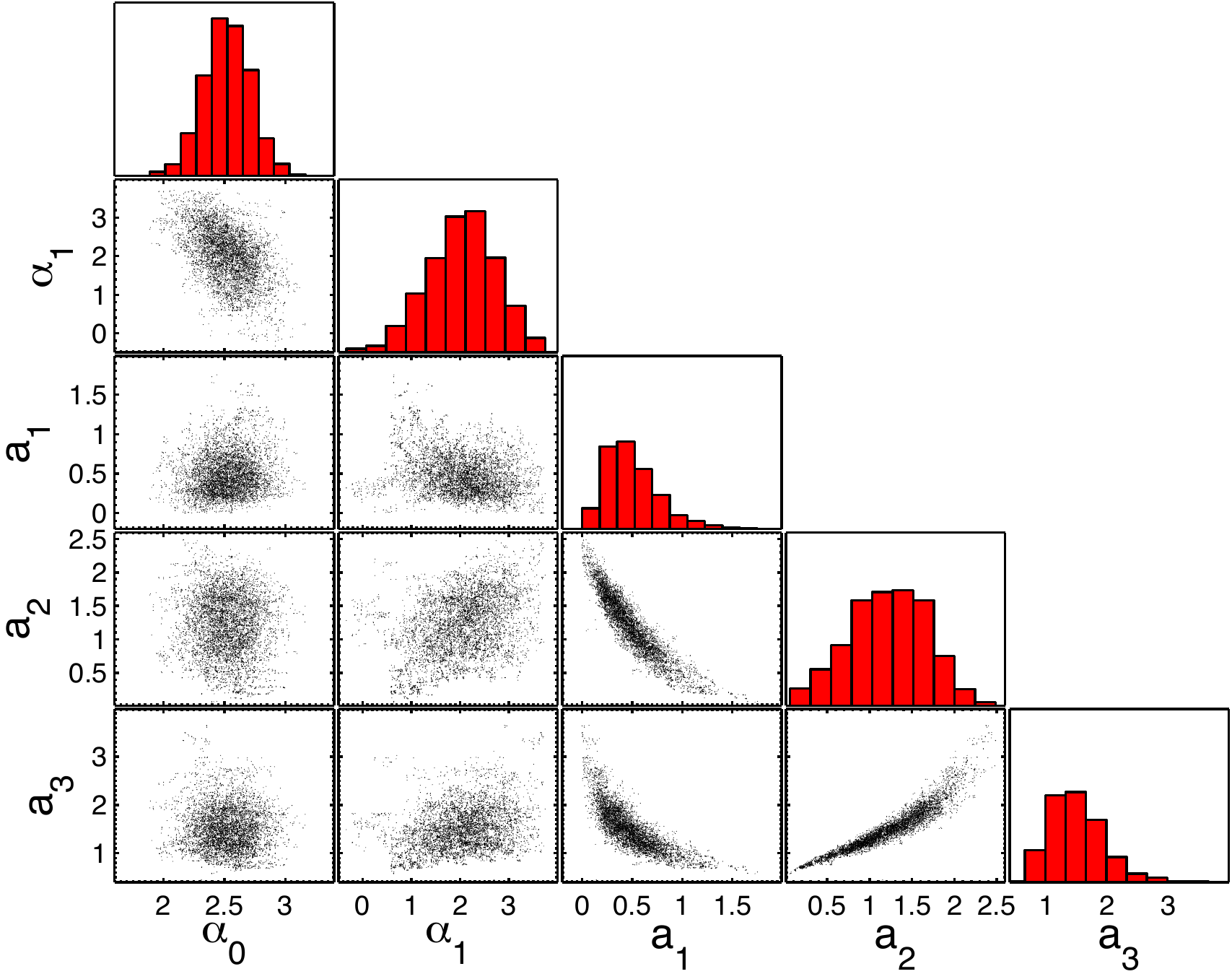}
\caption{Random samples drawn from   $P_{\rm tot}$ ($l_{\rm max}=100$) with the MCMC algorithm.  
 }
 \label{fig:smplt}
\end{figure}

The 1D marginalized $\Delta \chi^2$ curves  in figure \ref{fig:cdf} do not contain any information on the covariance between the 
parameters. We could obtain  confidence limits for 2 parameters from the full $P_{\rm tot}$ by implementing  a similar marginalization as done above for the one parameter case. Unfortunately, this 
 will be extremely CPU time consuming. 
For  a visual impression 
  of the actual covariance,
we resort to the Metropolis-Hastings Markov Chain Monte Carlo (MCMC) algorithm to generate 
  random samples of $\theta_i$ drawn from $P_{\rm tot}(\{\theta_{ i }\})$.    Figure \ref{fig:smplt} plots 7200 such random samples for the fiducial cut on $l$. 
There is a strong correlation between $a_2$ and $a_3$ as well as $a_1$ and $a_2$.
In agreement with figure \ref{fig:cdf}, the distribution of the individual parameters (except $\alpha_0$) as represented by the histograms deviates from 
normal especially for $a_1$, $a_2$ and $a_3$. From the MCMC sampling we 
get    $<\alpha_0>=  2.52$,  $<\alpha_1=>  2.03$, $<a_1>=   0.49$, $<a_2>=    1.21$ and
$<a_3>=    1.03$ with corresponding  rms values of     $0.19$,   $ 0.70$,    $0.26 $,   $0.45 $  and  $0.45$. These  mean are within $1\sigma$ deviations from the corresponding  ML estimates in Table 1. 

The full 
 error matrix $C_{ij}=<(\Delta \theta_i)(\Delta \theta_j)>$  where $\Delta \theta_i=\theta_i-\theta^{\rm ML}_i$, can be estimated  from the MCMC sampling but this requires 
 generating an unrealistically   large number of random sets. Hence, 
 we approximate $C_{ij}$ as the inverse of the  Hessian  
 $
 H_{ij}={\partial^2 P}/{\partial \theta_i \partial \theta_j} $  at   $\theta_i=\theta^{\rm ML}_i$. 
 The matrices $H_{ij}$ and $C_{ij}$ are  listed in Table 1. The second row in Table 1 
 gives the $1\sigma$ error,  $\sqrt{C_{ii}}$.  Since  $H_{ij}$ and $C_{ij}$ are
 symmetric matrices by construction the resulting error  is also symmetric.    
The quantity  $\sqrt{C_{ii}}$ represents the error when marginalization  over all other parameters is performed. Note that,  when the other parameters are fixed at their  ML values,
 the $1\sigma $ scatter   is $1/\sqrt{H_{ii}}$.

\begin{figure} 
\includegraphics[width=0.49\textwidth]{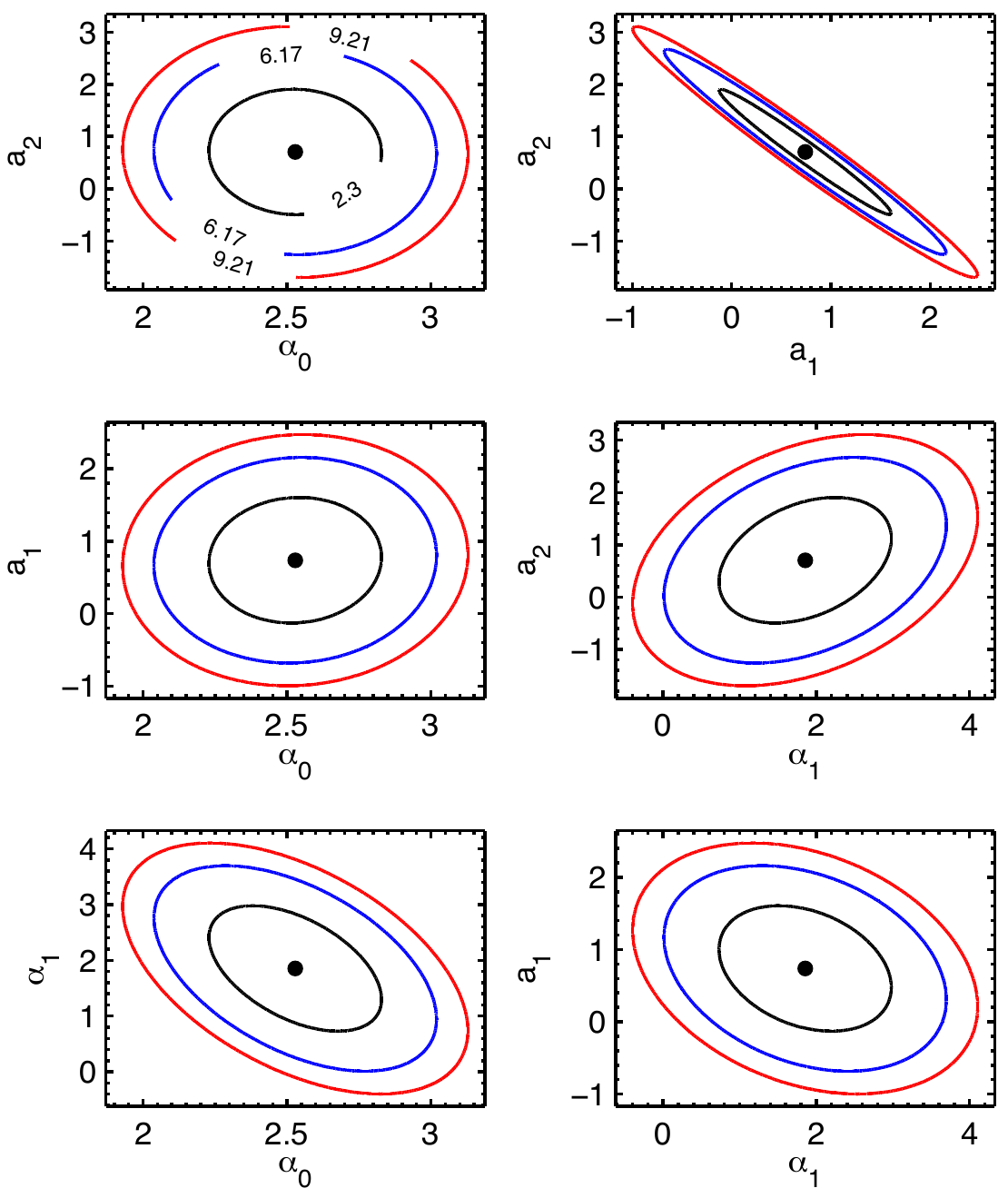}
\caption{
 Confidence levels in 2D projections of the parameter space. The contours $\chi^2=2.3$, 6.17, and $9.21$ correspond to $68\%$,  $95.4\%$ and $99\%$ confidence levels, respectively. In each panel, marginalization in performed over the remaining 3 parameters.}
 \label{fig:contouri}
\end{figure}

Confidence levels in 2D projections of  the parameter space are plotted 
in figure \ref{fig:contouri} as contours of  
$\Delta \chi^2=-2\ln(P_{\rm  tot}/P_{\rm max})\approx H_{ij}\Delta \theta_i\Delta \theta_j$. For each pair of parameters in each panel, $\Delta \chi^2$ is computed by marginalizing over the remaining parameters. 
For the sake of brevity,  we do not show projections involving $a_3$. This is justified since it is strongly correlated with $a_2$ and $a_1$ as seen   in figure \ref{fig:smplt}. The confidence levels in the  $\alpha_0-\alpha_1$ planel seem to agree with 
the distribution of points in the corresponding panel of figure \ref{fig:smplt}. The trend of decreasing $\alpha_0$ 
with increasing $\alpha_1$ is a reflection of the fact that a linear combination of these two parameters should be consistent with the same observed 
clustering amplitude. 

\begin{figure} 
\includegraphics[width=0.48\textwidth]{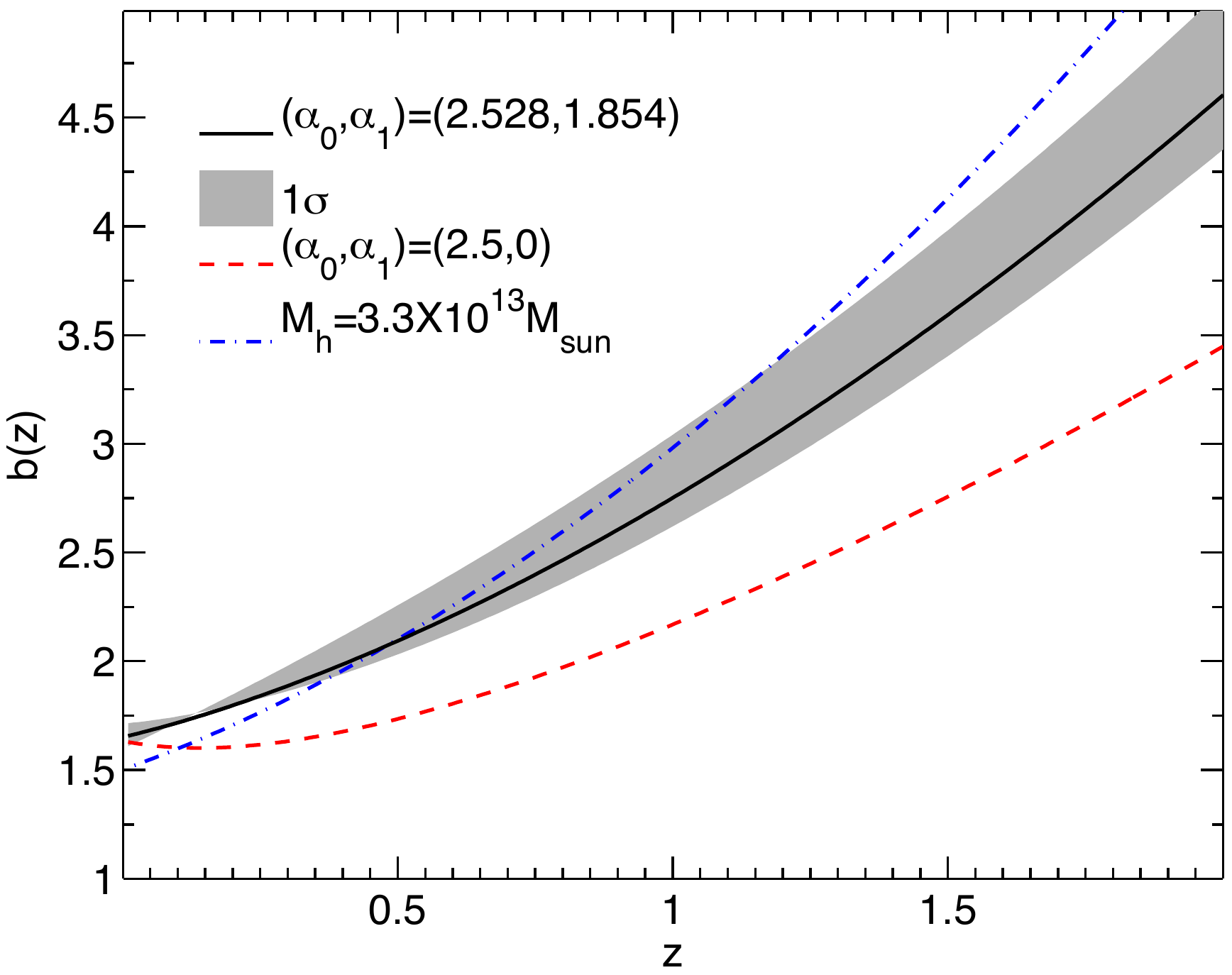}
\caption{The biasing factor of radio galaxies as a function of redshift.
Solid curve is obtained with the ML estimates for $\alpha_0$ and $\alpha_1$  and the shaded area represents $\pm 1\sigma$ deviation from the ML $\alpha_1$ (marginalization over $\alpha_0$ is performed).
For reference, the red dashed and blue dash-dot lines show, respectively,    the $b(z)$ 
for $\alpha_1=0$ and  for halos of mass $3.3\times 10^{13}M_\odot$. }
 \label{fig:bias}
\end{figure}

The bias factor (Eq.~\ref{eq:bias}) as a function of redshift  is plotted  in figure \ref{fig:bias}. The bias depends only on $\alpha_0$ and $\alpha_1$ and the black solid line
is computed for the ML values of these two parameters, as indicated in the figure.   
The shaded area  corresponds to   $\alpha_1$ taken within $\pm 1\sigma$  as computed from the marginalized $\Delta \chi^2$ in  the top-left panel in  figure \ref{fig:cdf}, where $\alpha_0$ is fixed at the maximum of $P_{\rm tot}$ for a given $\alpha_1$.
For comparison we also plot the case with no evolution in the radio activity versus stellar mass, i.e. $(\alpha_0,\alpha_1)=(2.5,0)$. 
 Halos of mass $3.3\times 10^{13}M_\odot$  (dot-dashed)  have a bias factor  close to the model prediction at $z=0.5$, but they are less clustered at lower redshifts. The halo mass corresponding to the model prediction of $b=1.66$ at $z=0$
 is $\sim 4.5\times 10^{13}M_\odot$.

\subsection{Dependence on $l$-cut and $M_{\rm min}$}
\begin{figure} 
\includegraphics[width=0.48\textwidth]{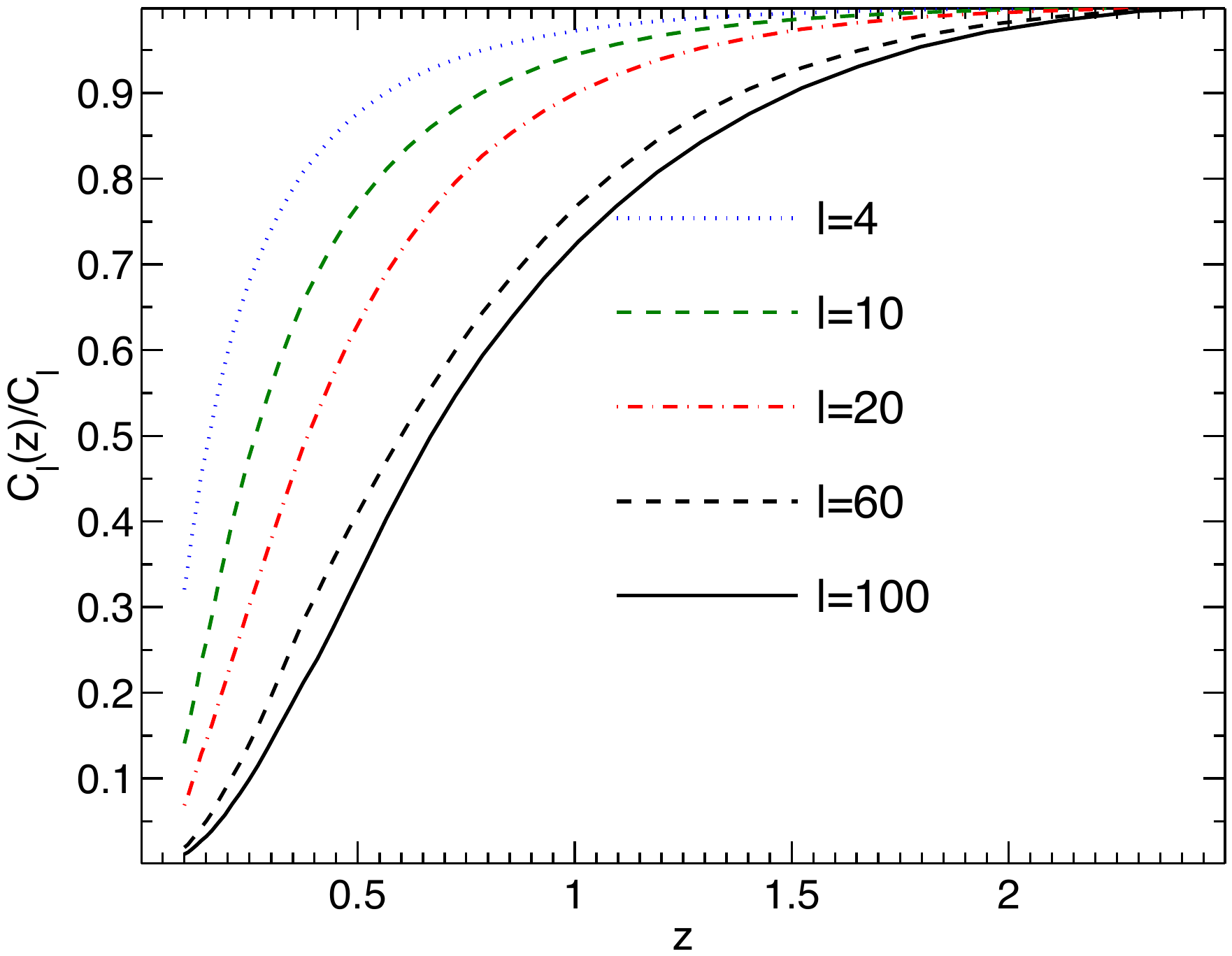}
\caption{The fractional contribution to $C_l$ resulting from integration from  redshift zero to $z$. The full $C_l$  is obtained with $z=3.5$. Only 5 values of $l$ presented, as indicated in the figure.  }
 \label{fig:Clz}
\end{figure}

\begin{figure} 
\includegraphics[width=0.48\textwidth]{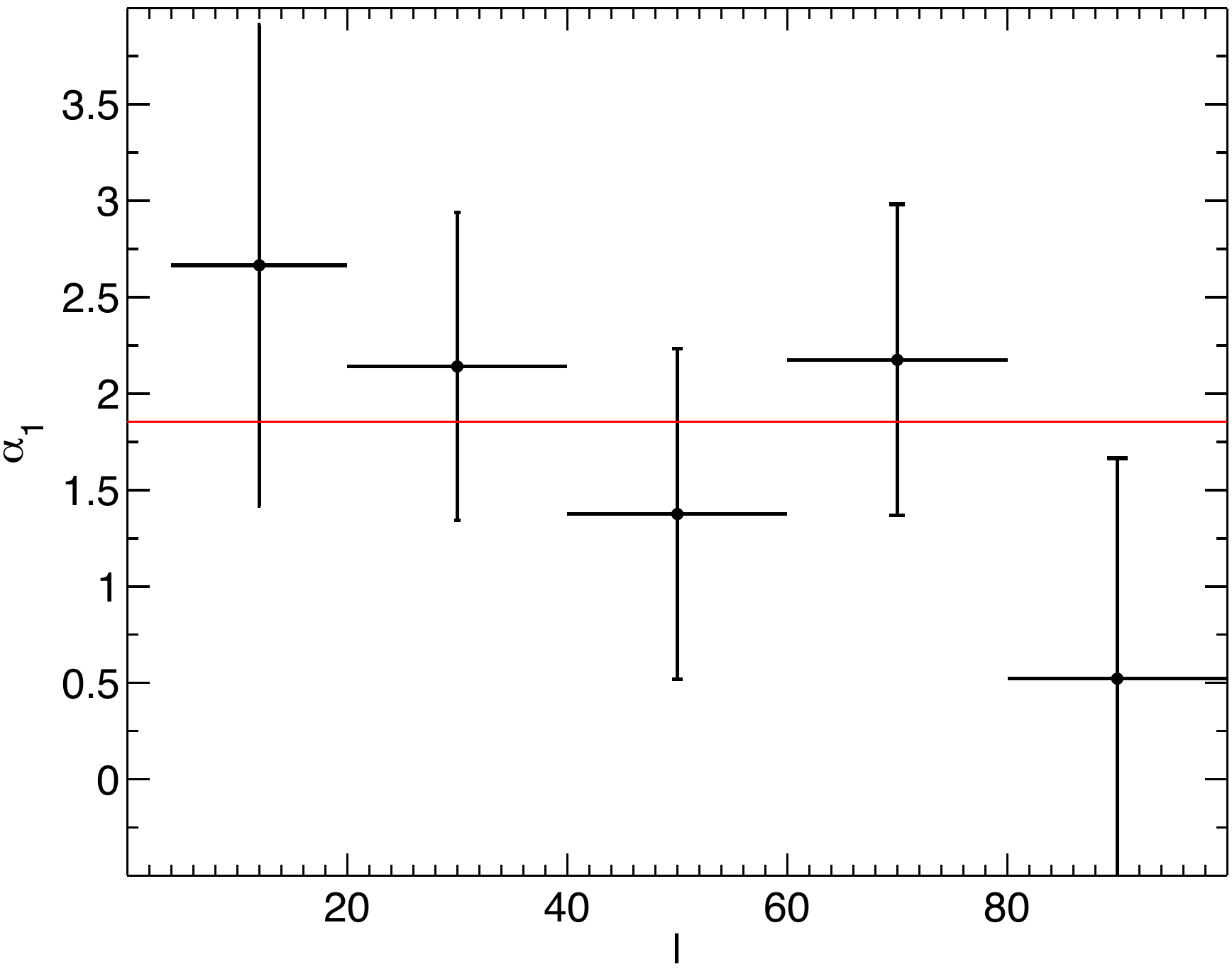}
\caption{ The inferred $\alpha_1$ from $C_l$ for different cuts on $l$. 
The horizontal bar spans  the  $l$ values used to obtain each point and  the error bar represents the  corresponding $1\sigma$ marginalized error.  }
 \label{fig:alphal}
\end{figure}

Different $l$ modes of the power spectrum probe  fluctuations at different depths (see e.g. Eq.~\ref{eq:limber}).  
In figure \ref{fig:Clz} we explore the relative contribution to ${\tilde C}_l$ from fluctuations between 
$r=0$ and  $r=r(z)$ i.e. with  the  upper limit in (\ref{eq:limber}) replaced with $r(z)$. 
The lower mode used in the ML, $l=l_{\rm min}=4$, picks up about $50\%$ of
its total power already by  $z\approx 0.1$, while $l=10$ by $z=0.25$. The modes between $l=20$ and $100$ gain 
half their power  
at redshifts between 0.35 and 0.7, with little difference between  $l=60$ and $100$.
In view of this, it is interesting to check the sensitivity of the ML estimate of  $\alpha_1$ to 
to different cuts imposed on  $l$.
Figure \ref{fig:alphal}  shows  ML estimates of $\alpha_1$
for  5 different cuts on $l$, i.e. on $l_{\rm min}$ and $l_{\rm max}$. Here the   remaining 4 model parameters are maintained  at the ML estimates  
obtained with the fiducial cut  $(l_{\rm min},l_{\rm max})=(4,100)$. 
The horizontal bars cover these 5  ranges in $l$ ($\Delta l=20$ except for the lowest point with $\Delta l=16$) and the vertical lines are 1$\sigma $ error bars. All estimates are consistent with each other  and with the ML value from the fiducial cut (horizontal red line) at less than the $1\sigma$ level.
There seems to be a hint for a  decreasing $\alpha_1$ for $l<60$,  but it is statistically insignificant.

\begin{figure} 
\includegraphics[width=0.48\textwidth]{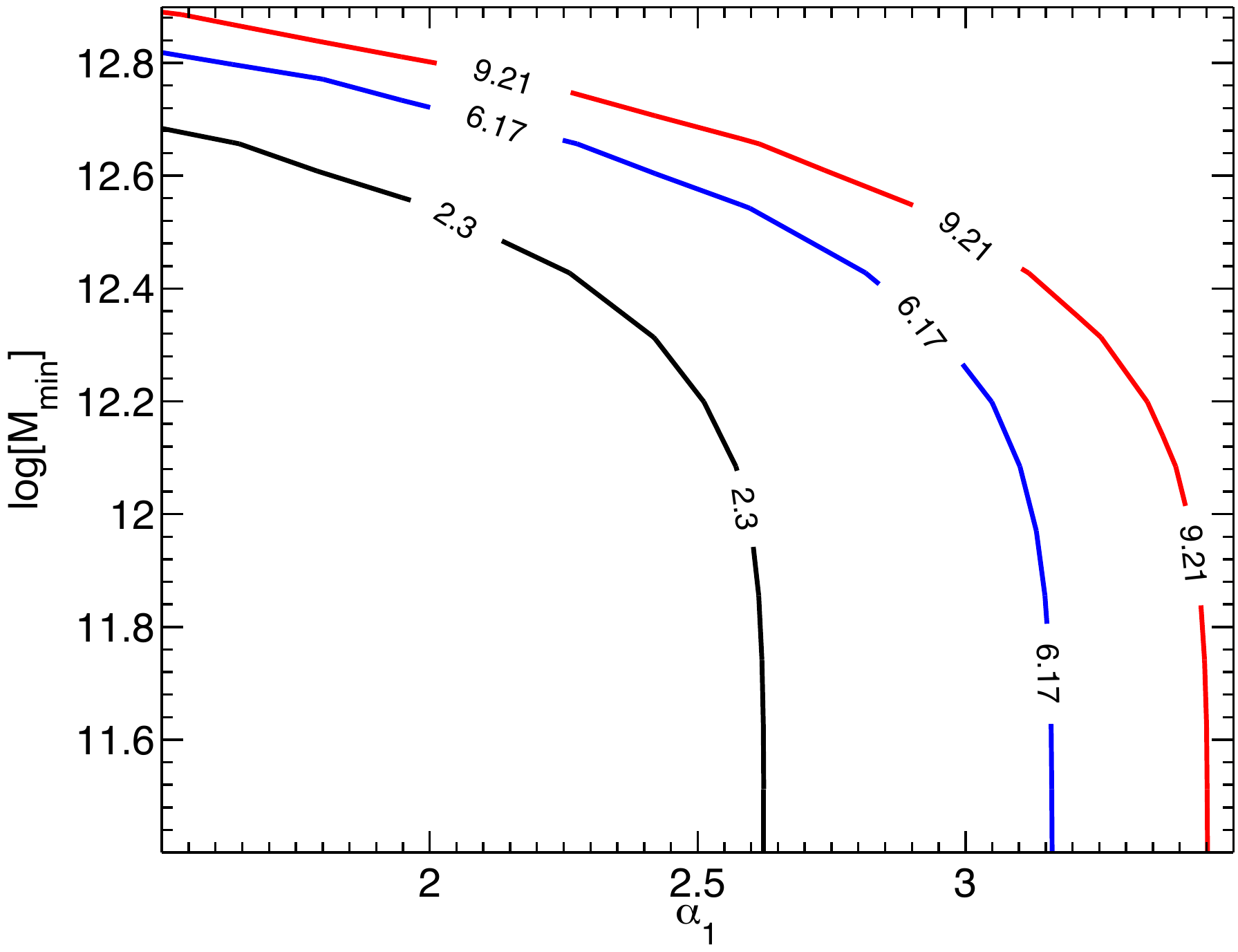}
\caption{ Confidence levels   in the plane of $\alpha_1$ and the minimum halo mass (in solar masses)
for hosting radio galaxies. }
 \label{fig:Mmin}
\end{figure}

 So far  we have assumed that radio activity exists only in halos with 
 mass above  $M_{\rm min}=4\times 10^{11}M_\odot$. We try now to check whether the clustering analysis could yield interesting constraints on $M_{\rm min}$. We fix the $N^{\rm model}$ parameters, $a_1$, $a_2$ and $a_3$ at the ML 
 values in Table 1 and compute $P_{\rm tot}$ in the plane of $\alpha_1$ and $M_{\rm min}$. Further, we marginalize over $\alpha_0$ as usual, i.e. for a given  point in the plane we use $\alpha_0$ which renders a minimum in $P_{\rm tot}$. 
In figure \ref{fig:Mmin} we plot  $\Delta \chi^2$, defined as  twice the negative of the logarithm of the ratio between $P_{\rm tot}$   evaluated at a particular point in the plane and the maximum value. 
The inner contour, corresponding to the 68\% confidence limit, encompasses a relatively large region, preventing  useful constraints on the threshold halo mass.
The bias factor, $b$, at $z=0.5$ changes very little for points inside this confidence level provided that $\alpha_0$ which 
renders a minimum in $P_{\rm tot}$ (i.e. the one obtained in the marginalization procedure) is used.  This is hardly surprising since $b$ has to be consistent with the observed angular power spectrum.



\begin{table*}
\caption{ML estimates  of the model 5 parameters and the corresponding error estimates for $l_{\rm max}=100$.
The Hessian (second derivatives of $-\ln P$ with respect to the fitting parameters at the MLE values) and 
the error covariance (inverse of the Hessian) are also listed. }
\centering
  \begin{tabular}{| c |c | c | c | c | c | c | }
    \hline
  &$\alpha_0$&$\alpha_1$ & $a_1$ & $a_2$ & $a_3$ \\ \hline
    &$2.529^{+0.184}_{-0.184}$       & $1.854^{+0.708}_{-0.761}   $   &$0.739^{+1.077}_{-0.382}$  &$0.705^{+0.689}_{-0.649}$  &  $1.057^{+0.550}_{-0.506}$\\
       & $\pm 0.197$     & $\pm 0.742$      &$\pm 0.572$      &$\pm 0.792$     & $\pm 0.531$\\
    \hline\hline
     \multicolumn{6}{c}{Hessian} \\\hline
         $\alpha_0$&54.138   &    13.431   &   -80.471  &    -149.24      & 132.57\\ \hline
         $\alpha_1$&13.431      &  6.453     & -34.429     & -66.526      &  60.32\\ \hline
         $a_1$      &-80.471     & -34.429    &   344.96     &  610.78    &  -534.96 \\ \hline
         $a_2$      &-149.24   &   -66.526    &   610.78     &  1203.2    &  -1125.2 \\ \hline
          $a_3$ &132.57      &  60.32    &  -534.96      &-1125.2   &    1093.8\\ \hline
\multicolumn{6}{c}{Error covariance} \\\hline
    $\alpha_0$ &0.039 &-0.073& 0.005 &-0.003 &-0.001\\ \hline
   $\alpha_1$&-0.073 &0.550 &-0.129 &0.204 &0.125 \\ \hline
    $a_1$ &0.005 &-0.129 &0.327 &-0.444& -0.290\\ \hline
   $a_2$ &-0.003 &0.204 &-0.444 &0.627& 0.417 \\ \hline
   $a_3$ &-0.001 &0.125 &-0.290 &0.417 &0.282\\ \hline
  \end{tabular}
  \end{table*}

\section{consistency with luminosity function }
\label{sec:consist}
\begin{figure}
\includegraphics[width=0.48\textwidth]{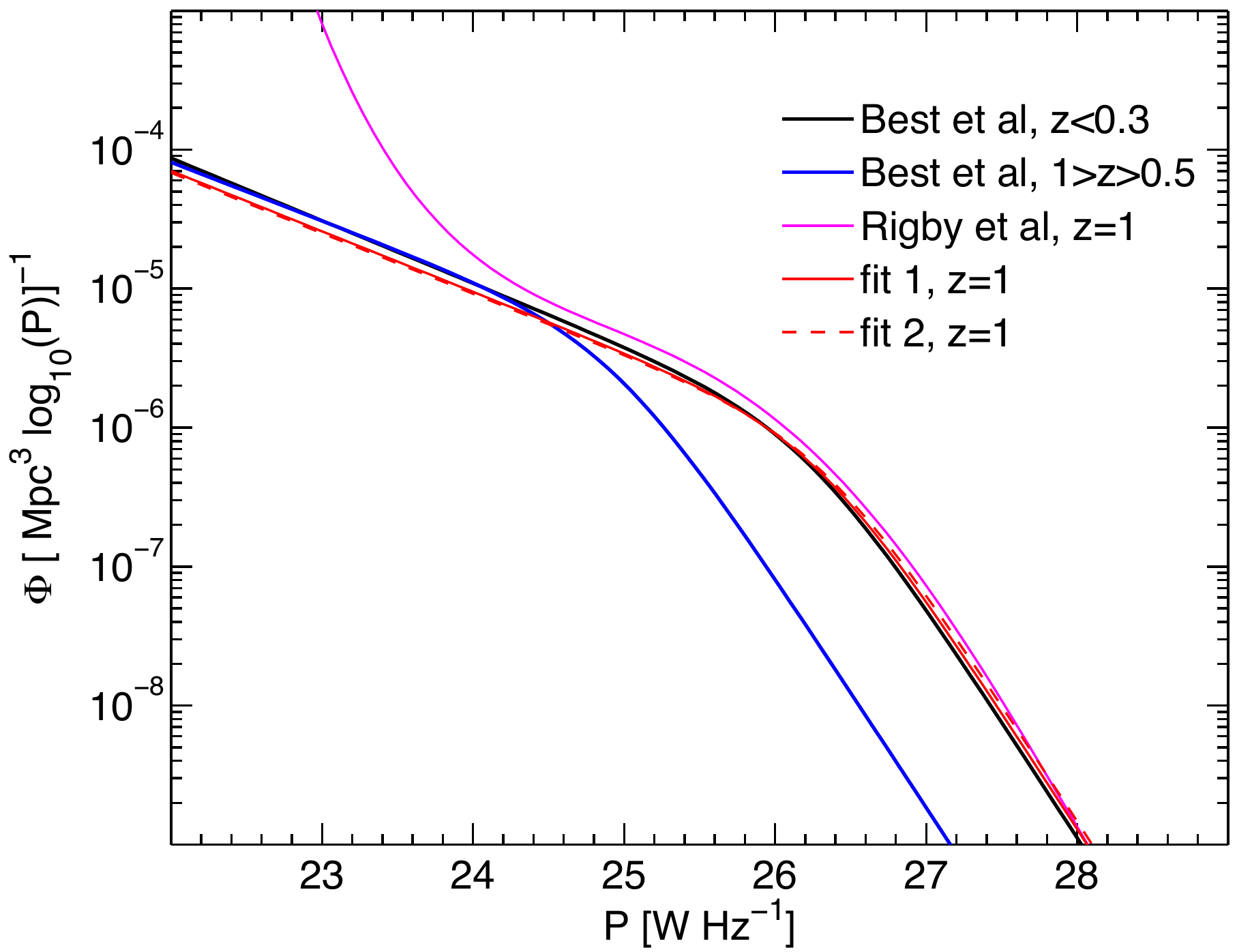} 
\caption{Various approximations to the luminosity functions. The red solid and dashed curves (fit 1 \& fit 2) correspond to the luminosity functions obtained as explained in the text.}
 \label{fig:phi}
\end{figure}

\begin{figure} 
\includegraphics[width=0.48\textwidth]{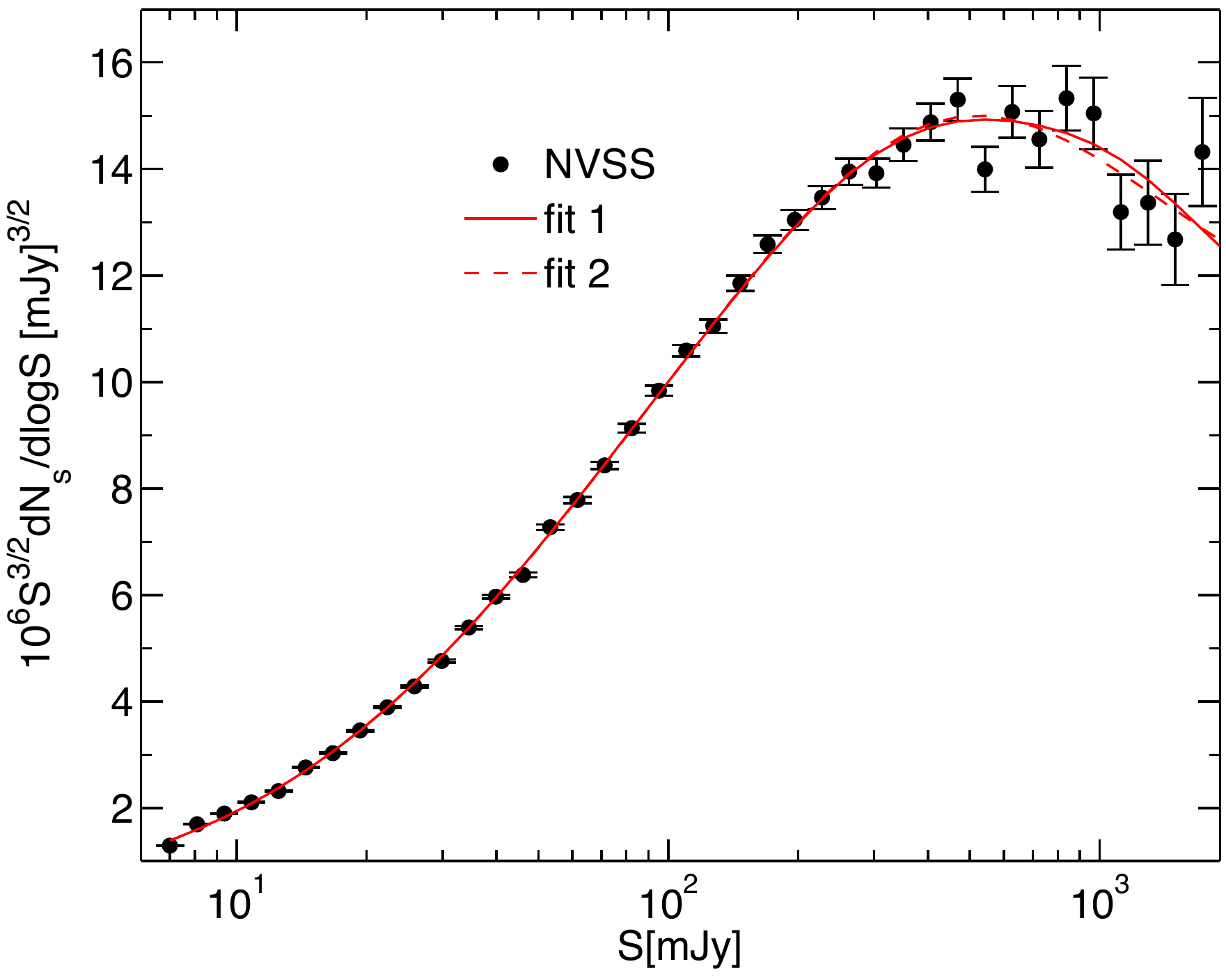}
\caption{Source count for the luminosity functions corresponding to fit \& fit 2 compared to the NVSS 
observations. }
 \label{fig:NS}
\end{figure}

The joint catalog of CENSORS and Hercules  used here for modeling $N(z)$ contains a small number of galaxies.
It is therefore interesting to  test whether our model $N^{\rm model}(z)$ is consistent with  a realistic luminosity function
defined as the number of sources per ${\rm log}(P)$ per comoving $\rm Mpc^3$.
We express the radio  luminosity function, $\phi(P,z)$,  as the sum of the star forming (SF) galaxies and AGNs luminosity functions \citep{Mauch2007}, $\phi_{_{\rm SF}}$ and $\phi_{_{\rm AGN}}$. 
For the SF, we follow \cite{Smolcic2008}  \citep[see also][]{Rigby2011} 
to write 
\begin{equation}
\phi_{_{\rm SF}}(P,z)=\phi_{_{\rm SF}}\left(\frac{P}{(1+z)^{2.5}},0\right)
\end{equation}  for $z\le z_{\rm max}=2.5$ and $P_*(z)=P(z_{\rm max})$ otherwise,  and we use the parametrized form in  \cite{Mauch2007} for  $\phi_{_{\rm SF}}$ at $z\approx 0$. In practice, the contribution of SF is small and the exact form of $\phi_{_{\rm SF}}$ makes little difference to our analysis here. 
For the AGN luminosity distribution we write
\begin{equation}
\label{eq:agn}
\phi_{_{\rm AGN}}(P,z)=\frac{C}{(P/P_*)^\alpha+(P/P_*)^\beta}\; ,
\end{equation}
where the  redshift dependence is via $C(z)$, $P_*(z)$, $\alpha(z)$ and $\beta(z)$ which  are   
expanded in  a Taylor series in $y=\ln(1+z)$.
The coefficients of this expansion are found by satisfying the following constraints 
\begin{itemize}
\item the observed  local $z\approx 0$ AGN luminosity function \citep[e.g.][]{Best2014a}
\item the NVSS source count, $N_s(S)$, giving the number of sources per  flux unit. 
\item the redshift distribution of radio  with $S>7.2 \rm mJy$ obtained from the 165 galaxies in Hercules and CENSORS.
\end{itemize} 
There is no physical or mathematical reasoning for fixing  the  order of the expansion for any of the redshift dependent functions in $\phi_{_{\rm AGN}}$. 
We adopt the expansions, 
\begin{eqnarray}
\nonumber
P_*(z)&=&P_0+p_1  y+p_2 y^2 +p_3 y^3 \\
\nonumber
C&=&C_0+p_4 y +p_5 y^2+p_6 y^3 \\
\nonumber \alpha&=&\alpha_0+p_7 y+p_8 y^2\\
\beta&=& \beta_0 \; .
\end{eqnarray}
Demanding that the form  (\ref{eq:agn}) reduces to 
the observed local luminosity function yields, $C=10^{-5.33} [{\rm log_{10} } P]^{-1} \rm Mpc^{-1}$, 
$P_0=10^{24.95} \rm W Hz^{-1}$,   $\alpha_0=1.66$
and $\beta_0=0.42$. The redshift dependence of the faint end slope, $\beta$, is  poorly probed by the source count, $N_s(S)$ and $N(z)$. 
Therefore, we simply take $\beta=\beta_0$.
 For a given choice for the parameters $p_i$, the luminosity yields the number of sources per unit redshift per unit 
 ${\rm log}(S)$
 \begin{equation}
 n(S,z)=4\pi D_c^2\frac{d D_c}{\dd z}\phi(P,z)\; ,
\end{equation}
 where $D_c$ is the comoving distance to redshift $z$ and $P=P(S,z)$ is the luminosity 
 corresponding to flux $S$ computed for a $\Lambda$CDM cosmology and a  power law spectral energy distribution,
 $\nu^{-0.8}$. 
 Predictions for the observed quantities $N(z)$ and $N_s(S) $ are  easily derived  from $n(S,z)$.
  The parameters $p_i$ are 
 then obtained by maximizing the likelihood for observing  these quantities. 
   Table  2 gives the parameter $p_i$ along with the Hessian and the error covariance matrices. In ``fit 1" in the table,
  galaxies with redshifts $1.45<z<2$ were excluded in the  $N(z)$, while ``fit 2" includes those galaxies.
  The AGN luminosity functions for these two fits computed at $z=1$ are shown in figure \ref{fig:phi} together with other 
  forms taken from \cite{Best2014a} and \cite{Rigby2011} (Fit 4 in their appendix D). There is a general agreement among all the curves for 
  $P>10^{25} \rm W Hz^{-1}$. 
   Figure \ref{fig:Nz} shows the predictions from the luminosity  functions derived here as the red solid (fit 1) and 
  red dashed (fit 2) curves. There is a very nice agreement with the observations and the other approximations to  $N(z)$
  plotted in the figure. An impressive agreement is also found between the predicted $N_s(S)$ and the observations as 
  seen in figure \ref{fig:NS}.

 \begin{table*}
   \caption{Parameters of luminosity function}
   \centering
  \begin{tabular}{|l |  l| l | l | l | l | l | l | l | }
    \hline
         & $p_1$  & $p_2$    & $p_3$  & $p_4$       & $p_5$     & $p_6$    & $p_7$      & $p_8$ \\ \hline
 Fit 1 &0.15  &  3.3   & -1.0 &    0.34   & -2.33    & 0.56   & -0.9    & 1.36 \\
          & $\pm 0.14$ &  $\pm  0.31$ &   $\pm 0.15$&  $\pm   0.11$    & $\pm 0.26$      &$\pm 0.13$     &$\pm 0.22$         &$\pm 0.32$\\  \hline
          Fit 2  & -0.183  &  4.32  & -1.635   & 0.74   &-3.381  &  1.146  & -0.445  &  0.691 \\
          & $\pm 0.32$ &  $\pm  0.84$ &   $\pm 0.46$&  $\pm  0.2$    & $\pm 0.53$      &$\pm 0.29$     &$\pm 0.24$         &$\pm 0.3$\\  \hline\hline
               \multicolumn{9}{c}{$10^{-6}$ Hessian (for fit 1 only)} \\\hline
 $p_1$    &0.987 &   0.836  &   0.806   &  1.315  &   1.168   &  1.175 &   -0.003  &   0.001\\  \hline
 $p_2$    &0.836 &    0.716  &   0.698  &   1.085  &   0.971  &   0.985 &   -0.012 &   -0.005\\  \hline
  $p_3$   &0.806  &   0.698  &   0.688  &   1.019  &   0.919   &  0.940 &   -0.020  &  -0.011\\  \hline
 $p_4$    &1.315   &  1.085  &   1.019  &   1.910  &   1.664   &  1.647   &  0.037  &   0.028\\  \hline
  $p_5$   &1.168  &   0.971   &  0.919  &   1.664   &  1.459   &  1.451   &  0.024  &   0.019\\  \hline
 $p_6$    &1.175  &   0.985  &   0.940   &  1.645  &   1.451   &  1.451   &  0.015   &  0.014\\  \hline
$p_7$    &-0.003  &  -0.012  &  -0.020   &  0.037  &   0.024   &  0.015  &   0.012   &  0.008\\  \hline
 $p_8$    &0.001  &  -0.005  &  -0.011  &   0.028  &   0.019   &  0.014   &  0.008  &   0.005\\  \hline\hline
     \multicolumn{9}{c}{ Error covariance (for fit 1 only)} \\\hline
  $p_1$    & 0.018   &-0.041 &   0.019 &  -0.013   & 0.028  & -0.013  & -0.004  & -0.002\\  \hline
   $p_2$    &-0.041  &  0.096 &  -0.047  &  0.027  & -0.065   & 0.032    &0.022   &-0.016\\  \hline
   $p_3$    & 0.019   &-0.047  &  0.024  & -0.012  &  0.031  & -0.016  & -0.016    &0.018\\  \hline
   $p_4$    &-0.013    &0.027  & -0.012  &  0.013   &-0.029 &   0.014  & -0.002   & 0.009\\  \hline
   $p_5$    & 0.028  & -0.065  &  0.031   &-0.029   & 0.068   &-0.034  & -0.005   &-0.007\\  \hline
   $p_6$    &-0.013  &  0.032  & -0.016   & 0.014  & -0.034   & 0.017  &  0.006   &-0.003\\  \hline
   $p_7$    &-0.004   & 0.022   &-0.016  & -0.002  & -0.005  &  0.006  &  0.047   &-0.067\\  \hline
   $p_8$    &-0.002  & -0.016   & 0.018  &  0.009  & -0.007   &-0.003 &  -0.067  &  0.106\\  \hline
          \end{tabular}
  \end{table*}

\section{Summary and Discussion}
\label{sec:summary}

We have found a mild evidence  ($\alpha_1>0$ at the 89\% confidence level) for evolution in the radio activity versus stellar mass relation. 
If all other parameters are fixed at their ML values, i.e. assuming that they are known we get a (symmetric) $1\sigma$  error 
$1/\sqrt{H_{22}}=0.51$. This is the best accuracy that the clustering of the NVSS can yield for $\alpha_1$ if $N(z)$ and $\alpha_0$ are known.
The constraint on $\alpha_1$ 
is obtained by using   observationally motivated SHM relation   \citep{Moster2013}.
We have not explored the effect of systematic uncertainties in this relation which in the relevant  redshift range ($z\sim 0.3-0.7$)
relies on a sample of  28000 sources selected with Spitzer  \citep{PerezGonzalez2008}.
Nonetheless, inspecting figure 10 in \cite{Coupon2015} 
shows that the \cite{Moster2013} relation which is adopted here is in very good agreement 
with the majority of observations plotted in this figure.
Nonetheless we cannot rule out the possibility that $\alpha_1>0$ may indicate that the 
\cite{Moster2013}  SHM ratio needs to be slightly modified at $z\sim 0.5$. 
Another  viable  interpretation for $\alpha_1>0 $,
is that halos hosting radio galaxies  could be more 
massive than    halos of other galaxies of the same stellar mass. This interpretation is in line with the conclusion  of  \cite{Mandelbaum2009}
based on small scale clustering and galaxy galaxy lensing.

The bias factor is less sensitive to the SHM 
 since it has to yield a theoretical $\cl$ which matches  the observed angular clustering within the $\Lambda$CDM model.
For  the ML parameters (Table 1) we find an excellent match between the theoretical and  observed $\cl$.
Further,  estimates of $\alpha_1$ obtained for different  cuts imposed on  $l$ 
are consistent with ML estimate obtained from the fiducial cut. The evidence for a dependence of 
$\alpha_1$ on scale (cf. Fig.~\ref{fig:alphal}) is statistically insignificant.

 \cite{Mandelbaum2009} examined the angular clustering  of 5700
 radio loud AGNs.  They   analyzed two point correlations at  projected separations 
  $\ltsim 10 \hmpc$ in the frame work of  
   halo occupation modeling \citep[e.g.][]{Peacock2000,White2001,Zheng2005} implemented  in the Millennium simulation. In contrast to our work which is restricted to large scales  ($\gtsim 50 \hmpc$), their modeling 
   depends on the division of radio galaxies as satellites and centrals. 
  From the clustering measurement alone they obtain a mean mass of $3\times 10^{13}M_\odot$, 
   for the central halos.
The bias factor we infer  (cf. Fig.~\ref{fig:bias})  
  matches that of  $3.3\times 10^{13}M_\odot $ halos, in agreement with  \cite{Mandelbaum2009}.
  It is reassuring  that the two independent studies, probing different 
  scales, yield similar halo  masses. 
However, incorporating galaxy-galaxy lensing in addition to clustering 
  they obtain halo masses 
   $(1.6\pm0.4) \times 10^{13}M_\odot$. \cite{Mandelbaum2009}
 argue that the deviation 
   could be attributed to uncertainties in the halo modeling, which mainly affects their clustering results.
   In general, it seems that galaxy-galaxy lensing incorporated in halo modeling tends to push halo masses to lower values, as revealed by inspecting  figure~10 in \cite{Leauthaud2012}.
   Nonetheless, our result is independent of the halo model. The cause of the  tension between the halo mass estimates
   based on lensing and clustering, remains obscure.  
  
\section{Acknowledgments}
We thank Curtis Saxton for a thorough reading of the manuscript.
This research was supported by
the I-CORE Program of the Planning and Budgeting
Committee, THE ISRAEL SCIENCE FOUNDATION
(grants No. 1829/12 and No. 203/09), the Asher Space
Research Institute, and the Munich Institute for Astro and
Particle Physics (MIAPP) of the DFG cluster of excellence
ÒOrigin and Structure of the UniverseÓ. 

\bibliography{Radio.bib,NVSS.bib}

\end{document}